\documentclass[runningheads]{llncs}
\usepackage{graphicx}
\usepackage{wrapfig}
\usepackage[colorinlistoftodos]{todonotes}

\usepackage{hyperref}
\usepackage[american]{babel}

\addto\extrasamerican{%
	\def\figurename{Figure}%
	\def\tablename{Table}%
}

\usepackage{booktabs}   %% For formal tables:
\usepackage{subcaption} %% For complex figures with subfigures/subcaptions
\usepackage{graphicx}
\usepackage{enumerate}
\usepackage{url}
\usepackage{amssymb}
\usepackage{pifont}

\usepackage{hyperref}
\usepackage{multirow}
\usepackage{tikz}
\usepackage{overpic}
\usepackage{ragged2e}

\begin{document}
\title{Self-Adaptive Digital Assistance Systems for Work 4.0}
\titlerunning{Self-Adaptive Digital Assistance Systems for Work 4.0}

\author{Enes Yigitbas\textsuperscript{[0000-0002-5967-833X]} \and Stefan Sauer\textsuperscript{[0000-0003-3084-0409]} \and \\ Gregor Engels\textsuperscript{[0000-0001-5397-9548]}}
\authorrunning{E. Yigitbas et al.}
\institute{Paderborn University\\ Zukunftsmeile 2, 33102 Paderborn, Germany\\
\email{firstname.lastname@upb.de}}
\maketitle              % typeset the header of the contribution
\begin{abstract}
In the era of digital transformation, new technological foundations and possibilities for collaboration, production as well as organization open up many opportunities to work differently in the future. The digitization of workflows results in new forms of working which is denoted by the term Work 4.0. In the context of Work 4.0, digital assistance systems play an important role as they give users additional situation-specific information about a workflow or a product via displays, mobile devices such as tablets and smartphones, or data glasses.

Furthermore, such digital assistance systems can be used to provide instructions and technical support in the working process as well as for training purposes. However, existing digital assistance systems are mostly created focusing on the “design for all” paradigm neglecting the situation-specific tasks, skills, preferences, or environments of an individual human worker. To overcome this issue, we present a monitoring and adaptation framework for supporting self-adaptive digital assistance systems for Work 4.0. Our framework supports context monitoring as well as UI adaptation for augmented (AR) and virtual reality (VR)-based digital assistance systems. The benefit of our framework is shown based on exemplary case studies from different domains, e.g. context-aware maintenance application in AR or warehouse management training in VR.   

\end{abstract}

\section{Introduction}

Nowadays we are witnessing a rising trend of digital transformation which is shaping our everyday life, value creation processes, and the way we are working. Especially in the context of production processes, the increasing amount of digitization and interconnection of production systems is sometimes referred to as \textit{Industry 4.0}. As a result of this industrial (r)evolution, the role of human work changes significantly, which is often denoted with the term \textit{Work 4.0} \cite{de2018work}. This means that in the context of Work 4.0 due to the digitization of workflows each individual worker will face a variety of challenges and problems to solve, mostly related to high cognitive activities. 

To overcome this problem, digital assistance systems play a crucial role to support human workers to execute their tasks in an efficient, effective, and pleasant manner. For this purpose, digital assistance systems assist human workers by providing additional situation-specific information about a workflow or a product via displays, mobile devices such as tablets and smartphones, or data glasses. Such digital assistance systems can be used to provide instructions and technical support in the working process as well as for training purposes. 

In the last decades, various digital assistance systems have been proposed in different application domains such as manufacturing~\cite{keller2019benefit}, assembly~\cite{hinrichsen2018digital}, or maintenance~\cite{DBLP:conf/ml4cps/KovacsAGUGS18}. 
However, existing digital assistance systems are created focusing on the “design for all” paradigm neglecting the situation-specific tasks, skills, preferences, or environments of an individual human worker. In most cases, the existing digital assistance systems are system-centred in a way that they primarily focus on the industrial task they are supporting. Certainly, in this connection, the context-of-use which is crucial for the interaction of the user with the production system is not considered. 

To overcome this issue, we present a monitoring and adaptation framework for supporting self-adaptive digital assistance systems (SADAS) for Work 4.0. According to Laddaga et al., a "Self-adaptive Software System evaluates its own behavior and changes behavior when
the evaluation indicates that it is not accomplishing what the software is intended
to do, or when better functionality or performance is possible"~\cite{laddaga2004self}. We make use of this definition and transfer the idea of self-adaptive software systems to self-adaptive digital assistance systems (SADAS). For this purpose, our framework supports context monitoring and UI adaptation for AR/VR-based SADAS. The benefit of our framework is shown based on example case studies from different domains, e.g. context-aware maintenance application in augmented reality or warehouse management training in virtual reality.   

The remainder of this book chapter is structured as follows: In~\autoref{sec:background}, we present background information on Industry\&Work 4.0 as well as digital assistance systems. In~\autoref{sec:challenges}, we discuss the main challenges in developing self-adaptive digital assistance systems. Based on these challenges, in~\autoref{sec:relWork}, we describe and discuss related approaches. ~\autoref{sec:framework} is dedicated to presenting our monitoring and adaptation framework for SADAS. In~\autoref{sec:caseStudies}, we present case studies to show the applicability of our framework. Finally,~\autoref{sec:conclusion} concludes our work with an outlook on future work.

\section{Background}
\label{sec:background}

In this section, we introduce basic concepts of \textit{Industry 4.0} and \textit{Work 4.0} as well as the main idea behind \textit{Digital Assistance Systems}.

\subsection{\textit{Industry} and \textit{Work 4.0}}

Since the beginning of industrialization, technological advancements have led to paradigm shifts which today are named "industrial revolutions": in the
field of mechanization (the so-called 1st industrial revolution), of the intensive
use of electrical energy (the so-called 2nd industrial revolution), and of the
widespread digitization (the so-called 3rd industrial revolution)~\cite{lasi2014industry}. On the basis of an advanced digitization within factories, the combination of internet technologies and future-oriented technologies in the field of “smart” objects (machines and products) the
term “Industry 4.0” was established for a planned “4th industrial revolution”~\cite{lasi2014industry}. 

According to~\cite{russmann2015industry}, the term \textit{Industry 4.0} stands for the fourth industrial revolution which is defined as a new level of organization and control over the entire value chain of the life cycle of products. For realizing the future of productivity and growth in manufacturing industries, \textit{Industry 4.0} includes several enabling technologies such as cyber physical systems (CPS), internet of things (IoT), cloud computing, or novel forms of human-computer interaction. 
Over the last few years, Industry 4.0 has emerged as a promising technology framework used for integrating and extending manufacturing processes at both intra- and inter-organizational levels. This emergence of Industry 4.0 has been fuelled by the recent development in ICT. The developments and the technological advances in Industry 4.0 provide a viable array of solutions to the growing needs of digitization in manufacturing industries~\cite{xu2018industry}. 

On the other hand, the process of digitization and incorporation of new technologies and intelligent systems in various sectors and domains, are the core enablers for the changes which are about to come with the new way of work~\cite{bonekamp2015consequences}. Nowadays, the business processes of every corporation and organization are supported by powerful IT systems which become more enhanced by the introduction of sophisticated robotic and sensor technologies, Cyber-Physical Systems, 3D printing technologies and intelligent software systems. As a consequence of the process of rapid digitization and the technology fluctuations, the requirements and demands for the working individuals in the workplace are changing. 

Therefore, the term \textit{Work 4.0} was introduced in November 2015 by the German Federal Ministry of Labour and Social Affairs (BMAS) when it launched a report entitled Re-Imagining Work: Green Paper Work 4.0~\cite{salimi2015work}. This initiative envisions new ways of work where the focus will be on the human workers, taking into account their individual abilities, characteristics, and preferences while aiming at allowing greater flexibility and ensuring work-life balance. Considering the current predictions, it becomes necessary to focus on the human as an important part in the sector of industrial production. Therefore, the need to develop digital assistance systems which are able to adapt to the personal abilities, needs, and individual characteristics of the working individuals is emerging.

\subsection{Digital Assistance Systems}

The term \textit{Digital Assistance System} was introduced in~\cite{hold2017planning} as the primary interface to optimally integrate humans into a production environment during task execution. Based on this definition, a DAS can be seen as a technical system for dynamically delivering digitally prepared information. Informational assistance systems record data via sensors and inputs, then process this data to provide employees the right information (“what”) at the right time (“when”) in the desired format (“how”) \cite{nikolenko2019digital}. The main goals of a DAS are to avoid uncertainty and mental stress for users, warn them of dangers, as well as increase of productivity, e.g., reduction of training time, search times, or operating errors~\cite{hold2017planning}. DAS can be divided into stationary assistance
systems, mobile assistance systems, handheld devices (such as tablet PCs), and wearables \cite{nikolenko2019digital}. While stationary assistance systems are permanently installed at a work station (such as a mounted projection device), mobile assistance systems, in contrast, are moved to the assembly object via a mobile solution. Wearables can be classified by the body part on which they are worn (such as “smart glasses,” “smart gloves,” “smart watches”). 

In the past, DAS have often been used to create standardized instruction manuals to be used by all employees working on the assembly system (design for all) – independent of their individual features. To tackle this limitation, providing personalized and situation-specific assistance is a promising alternative to empower the workers while supporting them in performing complex physical and cognitive tasks. However, in order to provide such assistance, the DAS needs to be enriched with capabilities concerning continuous monitoring and self-adaptation which are known from the area of \textit{Autonomic Computing}. The term Autonomic Computing (also known as AC) refers to the self-managing characteristics
of distributed computing resources, adapting to unpredictable changes while hiding
intrinsic complexity to operators and users~\cite{DBLP:journals/computer/KephartC03}. The AC system concept is designed to monitor and adapt a \textit{Managed Element}, using high-level policies.
It will constantly check and optimize its status and automatically adapt itself to changing conditions by using the Monitor, Analyze, Plan, Execute-Knowledge loop. Based on the ideas of context-awareness and self-adaptation, we aim to bring classical DAS to a new level of Self-adaptive Digital Assistance Systems (SADAS).

\section{Challenges}
\label{sec:challenges}

In the course of two different research projects, we have analyzed the challenges in the application and adoption of digital assistance systems in the industry setting. The first project was related to the manual assembly of an Electrical Cabinet (E-cabinet),
while the second one was dealing with the process of manual assembly of a concrete
product in a Smart Factory \cite{DBLP:conf/hci/JosifovskaYE19}. For identifying the requirements and needs of the human workers in the industrial sector with regard to the usage of DAS, we have conducted semi-structured interviews with experts from different research fields: Psychology, Sociology, Didactic, Economics, Computer Science, Electrical
and Mechanical Engineering. Based on this investigation, we have identified the following main challenges: \\

\textit{Challenge 1: Information Presentation}

For many years traditional graphical user interfaces (GUIs) have been successfully adopted for mobile platforms. e.g. through the integration of multi-touch interaction and responsive layout algorithms that adapt the visual display to different device sizes. However, in applications that rely on spatial information related to a real-world environment GUIs are not ideal, because the information displayed in the interface is removed from its real-world context and interaction is effected indirectly through the interface~\cite{DBLP:conf/etfa/Paelke14}. Especially digital assistance systems in the context of manufacturing and assembly using current sensor data and the user's current location are examples that rely heavily on such spatial information which can be reached through interaction technologies like \textit{Augmented Reality} or \textit{Virtual Reality}. Associated with the aspect of information presentation is the question of the computing platform how a digital assistance system works and can be accessed by the end-users. There are several target devices for DAS on the market which are developed by different companies and organizations. Target devices could be smartphones, tablets, or HMDs for Augmented Reality (e.g. Microsoft Hololens, RealWear Glasses, or Google Glass Enterprise Edition) or Virtual Reality (e.g. HTC Vive, Oculus Quest, or Valve Index). The cost of a device, its comfort in using it, and the ability of a device to help the user accomplish her task are some of the reasons that influence the type of equipment that different users and organizations use to acquire them. Each computing platform can have different properties regarding hardware and sensor, operating system, used SDKs, etc. Given the heterogeneous span of various devices for DAS, it is essential to have multi-platform support so that a digital assistance system can be deployed and used across varying computing platforms. 

\textit{Challenge 2: Monitoring}

The acceptance and successful application of a digital assistance system highly depends on the quality of the information that is shown to the end-users in guiding them through their task. With this regard, it is important to provide situation-aware information for the end-users so that they can accomplish their tasks in an efficient, effective, and satisfying manner. For this purpose, a digital assistance system (DAS) should enable context monitoring features to the end-users to inform them about dynamically changing characteristics of the working environment. With this regard, an important challenge is to continuously observe the context-of-use of a DAS through various sensors. The context-of-use can be described through different characteristics regarding the user (physical, emotional, preferences, etc.), platform (Hololens, Handheld, etc.), and environment (real vs. virtual environmental information). Due to the rich context dimension which is spanning over the real-world and virtual objects, it is a complex task to track and relate the relevant context information to each other. The mixture of real (position, posture, emotion, etc.) and virtual (coordinates, view angle, walk-through, etc.) context information additionally increases the aspect of context management compared to classical context-aware applications like in the web or mobile context. \\

\textit{Challenge 3: Adaptation}

Based on the collected context information, a decision-making process is required to analyze and decide whether conditions and constraints are fulfilled to trigger specific adaptation operations on the DAS. 
In general, an important challenge is to cope with conflicting adaptation rules which aim at different adaptation goals.
This problem is even more emphasized in the case of AR-based digital assistance systems as we need to ensure a consistent display between the real-world entities and virtual overlay information.
For the decision-making step, it is also important to decide about a reasoning technique like rule-based or learning-based to provide a performant and scalable solution. 

As AR/VR-based digital assistance systems consist of a complex structure and composition, an extremely high number of various adaptations is possible.
The adaptations should cover text, symbols, 2D images, and videos, as well as 3D models and animations.
In this regard, many adaptation combinations and modality changes increase the complexity of the adaptation process.

\section{Related Work}
\label{sec:relWork}

In previous work, different approaches were introduced to address the above-mentioned challenges \textit{Information Presentation}, \textit{Monitoring}, and \textit{Adaptation} within the scope of digital assistance systems.

In \cite{DBLP:conf/cdmake/FellmannRBMR17}, the authors present a framework for assistance systems to support work processes in smart factories. They argue that, due to to the large spectrum of assistance systems, it is hard to acquire an overview and to select an adequate digital assistance system based on meaningful criteria. Therefore, they suggest a set of comparison criteria in order to ease the selection of an adequate assistance system. Compared to our framework, this work is rather supporting the process of selecting a suitable digital assistance system while the above-mentioned challenges are not explicitly covered.

Similar work is presented in \cite{DBLP:conf/indin/GoreckySLZ14} where the authors present solution ideas for the technological assistance of workers. Besides technological means for supporting human-machine interaction in the Industry 4.0 era, the authors describe how the novel role of human workers in the context of Industry 4.0 should be addressed. As concrete examples for digital assistance systems, they focus on web-based and mobile apps incorporating AR features for Work 4.0 scenarios. They also focus on the aspects of context monitoring and adaptive UIs for the AR-based assistance app. However, the main focus is on hand-held AR devices, while the usage of head-mounted displays in AR or VR is not covered. 

A more formal approach in guiding different stakeholders to choose an adequate digital assistance system for their organization, domain, or application field is presented in \cite{DBLP:conf/ml4cps/KovacsAGUGS18}. This work proposes a process-based model to facilitate the selection of suitable DAS for supporting maintenance operations in manufacturing industries. Using this approach, a digital assistance system is selected and linked to maintenance activities. Furthermore, they collect user feedback by employing the selected DAS to improve the quality of recommendations and to identify the strength and weaknesses of each DAS in association with the maintenance tasks. While this approach supports the selection of an adequate DAS in the context of Industry 4.0 it is not focusing on the aspects of monitoring and adaptation. 

Besides the above-described approaches, some other works in the field of digital assistance systems apply the human-centered design process in order to design and develop assistance systems that fulfill the needs of the user requirements and the context-of-use. An example work in this direction is presented in \cite{DBLP:conf/icit2/NellesKMS16} where the authors develop a digital assistance system for production planning and control. Similar to our work, this approach is focusing on the aspects of context monitoring and UI adaptations within DAS, however, they do not apply AR/VR interfaces to cover spatial information and interaction with a DAS. 

Another type of work related to digital assistance systems is presented in \cite{DBLP:conf/hci/FischerSR018} where the authors propose a lightweight canvas method to foster interdisciplinary discussions on DAS. While this approach is primarily focusing on interdisciplinary discussions in the early stages of requirements understanding and design of DAS, it is not addressing a development process for situation-aware digital assistance systems in AR or VR. 

The most related approach to our monitoring and adaptation framework for SADAS is presented in \cite{DBLP:conf/hci/JosifovskaYE19}. In this work, the authors introduce a digital-twin based multi-modal UI adaptation framework for assistance systems in Industry 4.0. This approach characterizes a predecessor solution of our presented work here. While this work covers aspects of context-awareness and adaptation for DAS, the scope of targeted applications and devices remains restricted so that AR and VR technologies for example are not covered. 

While the above-described approaches highly focus on the selection and development of digital assistance systems, they are not fully covering the novel aspects of context-awareness and adaptation especially in the combination of AR and VR applied for DAS. Therefore, in the following, according to the challenges introduced in~\autoref{sec:challenges}, we analyze further approaches which focus more on the topic of context monitoring and adaptation within the scope of AR and VR applications.

Augmented Reality (AR) and Virtual Reality (VR) have been a topic of intense research in the last decades. In the past few years, massive advances in affordable consumer hardware and accessible software frameworks are now bringing AR and VR to the masses.  AR enables the augmentation of real-world physical objects with virtual elements and has been already applied for different aspects such as robot programming \cite{DBLP:conf/interact/YigitbasJE21}, product configuration (e.g., \cite{DBLP:conf/hcse/GottschalkYSE20}, \cite{DBLP:conf/hcse/GottschalkYSE20a}), prototyping \cite{DBLP:conf/hcse/JovanovikjY0E20}, planning and measurements \cite{DBLP:conf/eics/EnesScaffolding} or for realizing smart interfaces (e.g., \cite{DBLP:conf/eics/KringsYJ0E20}, \cite{DBLP:conf/interact/YigitbasJ0E19}). In contrast to AR, VR interfaces support the interaction in an immersive computer-generated 3D world and have been used in different application domains such as training~\cite{DBLP:conf/vrst/YigitbasJSE20}, prototyping \cite{DBLP:journals/corr/abs-2107-00377}, robotics \cite{DBLP:conf/seams/YigitbasKJE21}, education~\cite{DBLP:conf/mc/YigitbasTE20}, healthcare~\cite{DBLP:conf/mc/YigitbasHE19}, or even for collaborative software modeling \cite{DBLP:journals/corr/abs-2107-12772}. 

While context-awareness has been exploited in various types of applications including web \cite{YigitbasS0E17}, mobile (e.g., \cite{YigitbasJJKAE19} or \cite{DBLP:journals/pacmhci/YigitbasHRASE19}), and cross-channel applications (e.g., \cite{Yigitbas016} or \cite{DBLP:conf/hcse/YigitbasAJK0E18}) to improve the usability of an interactive system by adapting its user interface, only a few existing works are focusing on the topic of context-awareness in AR and VR.

In~\cite{GrubertLZR17}, the concept of Pervasive Augmented Reality (PAR) is introduced.
A taxonomy for PAR and context-aware AR that classifies context sources and targets is presented.
The context sources are classified as human, environmental, and system factors.
As apparent in the title, Grubert’s work treats Augmented Reality, here with special regards to pervasive Augmented Reality.

Context-aware Mobile Augmented Reality (CAMAR)~\cite{camar} is an approach on context-awareness in mobile AR focusing on user context, which is measured using the user’s mobile device.
It enables the user to customize the presentation of virtual content and to share this information with other users selectively, depending on the context.
Furthermore, a framework called UCAM (Unified Context-aware Application Model)~\cite{hong2006new} can be used to create CAMAR-enabled applications.
UCAM is a framework which besides the acquisition, process, and awareness of contextual information provides also a unified way of representation with respect to user, content, and environment.

The framework presented in~\cite{LindlbauerFH19} focuses on the context-aware adaptation of interfaces in mixed reality, with the main adaptation points being what content is displayed, where it is shown and how much information of it is displayed. It is designed to adjust the content display depending on the user’s tasks and their cognitive load and archives this using a combination of rule-based decisions and combinatorial optimization. The framework uses parameters about the applications that are to be displayed as input additionally to the context-specific parameters to achieve a fitting layout optimization. The framework is mentioning mixed reality as its base, but regarding that, it shows contents in the real world and does not create a whole new virtual world, it can safely be said that AR is supported. 

Apart from the above-mentioned approaches which address the development of context-aware applications in general without directly focusing on digital assistance systems in the context of Industry 4.0, there are also specific approaches that use AR in the smart factory context. One example of such an approach is presented in \cite{DBLP:conf/etfa/Paelke14}. Here, AR is used for supporting workers in an Industry 4.0 environment where they have to accomplish assembly tasks. The work presents the initial experience with the AR-based assistance systems.

\section{Monitoring and Adaptation Framework}
\label{sec:framework}

\begin{figure}[h]
  \centering
	\includegraphics[width=1\textwidth]{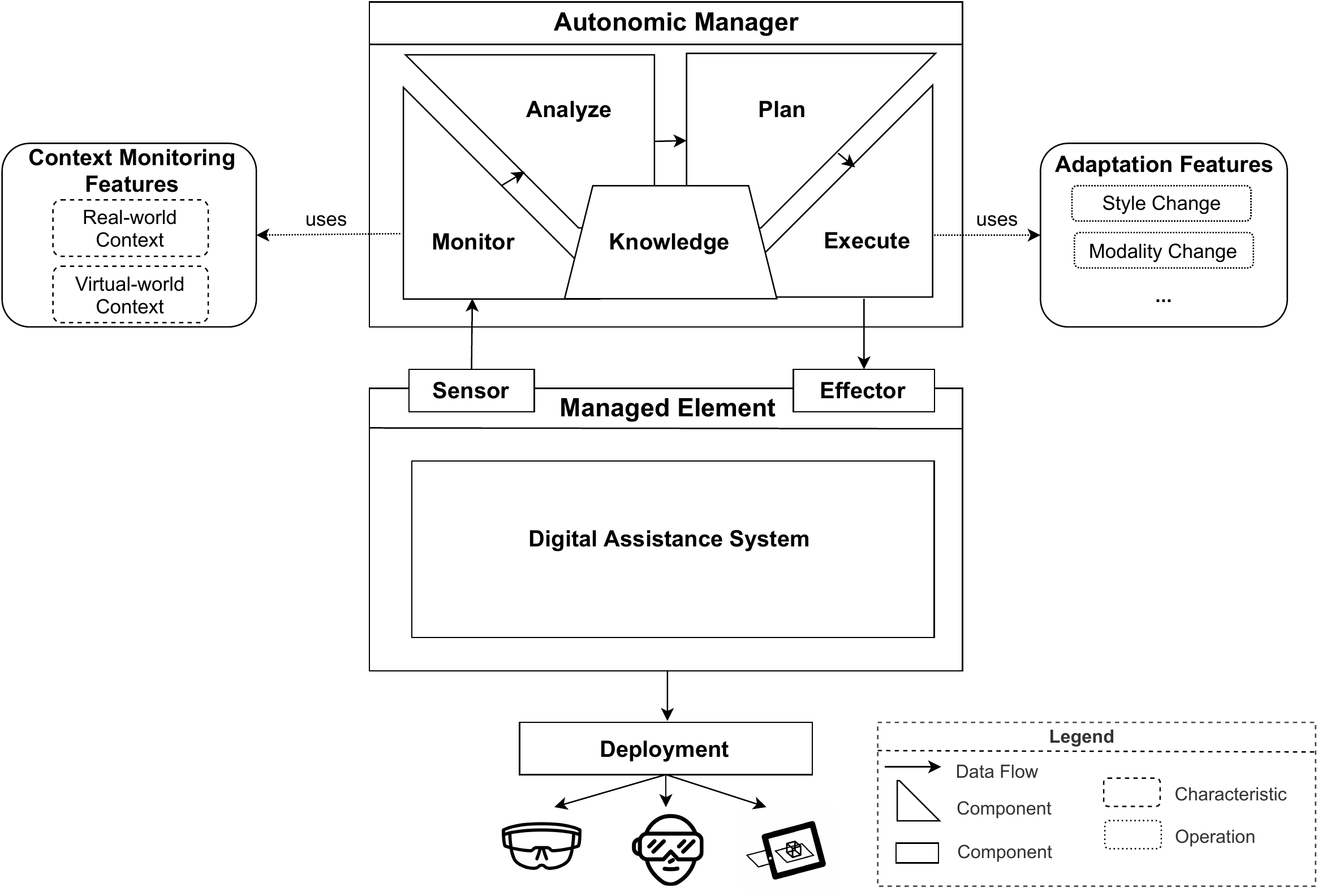}
    \caption{Architectural overview of the monitoring and adaptation framework}
  	\label{fig:intFramework}
\end{figure}

In order to address the described challenges, we present a monitoring and adaptation framework for supporting self-adaptive digital assistance systems (SADAS). Our framework which is based on the \textit{MAPE-K} architecture~\cite{DBLP:journals/computer/KephartC03}, is depicted in~\autoref{fig:intFramework}.

It is basically divided up into two main components, the \textit{Autonomic Manager} and the \textit{Managed Element}. The \textit{Autonomic Manager} is responsible for continuously monitoring the \textit{Managed Element} through \textit{Sensors} and to automatically react to changing conditions by adapting the \textit{Managed Element} through \textit{Effectors}. For this purpose, the \textit{Autonomic Manager} consists of a control loop that is called MAPE-K, while this acronym represents the starting letters of the main sub-components: \textit{Monitor}, \textit{Analyze}, \textit{Plan}, \textit{Execute}, and \textit{Knowledge}.

The \textit{Managed Element} represents in our case the \textit{Digital Assistance System (DAS)} that is deployed on an execution platform that can be accessed through different devices such as VR HMDs, AR Smart Glasses, or Tablets. Furthermore, the DAS is characterized through context information that can be observed through \textit{Context Monitoring Features}. This context information can consist either of \textit{Real-world Context} information which is gathered by sensing existing sensors in the real physical world (in the case of AR) or \textit{Virtual-world Context} information when context information such as gestures, pose, or virtual environment information are continuously monitored in the VR world. Besides context information that can be observed through the sensors of the DAS, there are \textit{Adaptation Features} to characterize the adaptation operations which are executed with the means of the \textit{Effectors} of the DAS. The \textit{Adaptation Features} can contain various adaptation operations to adjust the DAS interface through run-time adaptations, e.g., changing modality or layout.  

\begin{figure}[h]
  \centering
	\includegraphics[width=1\textwidth]{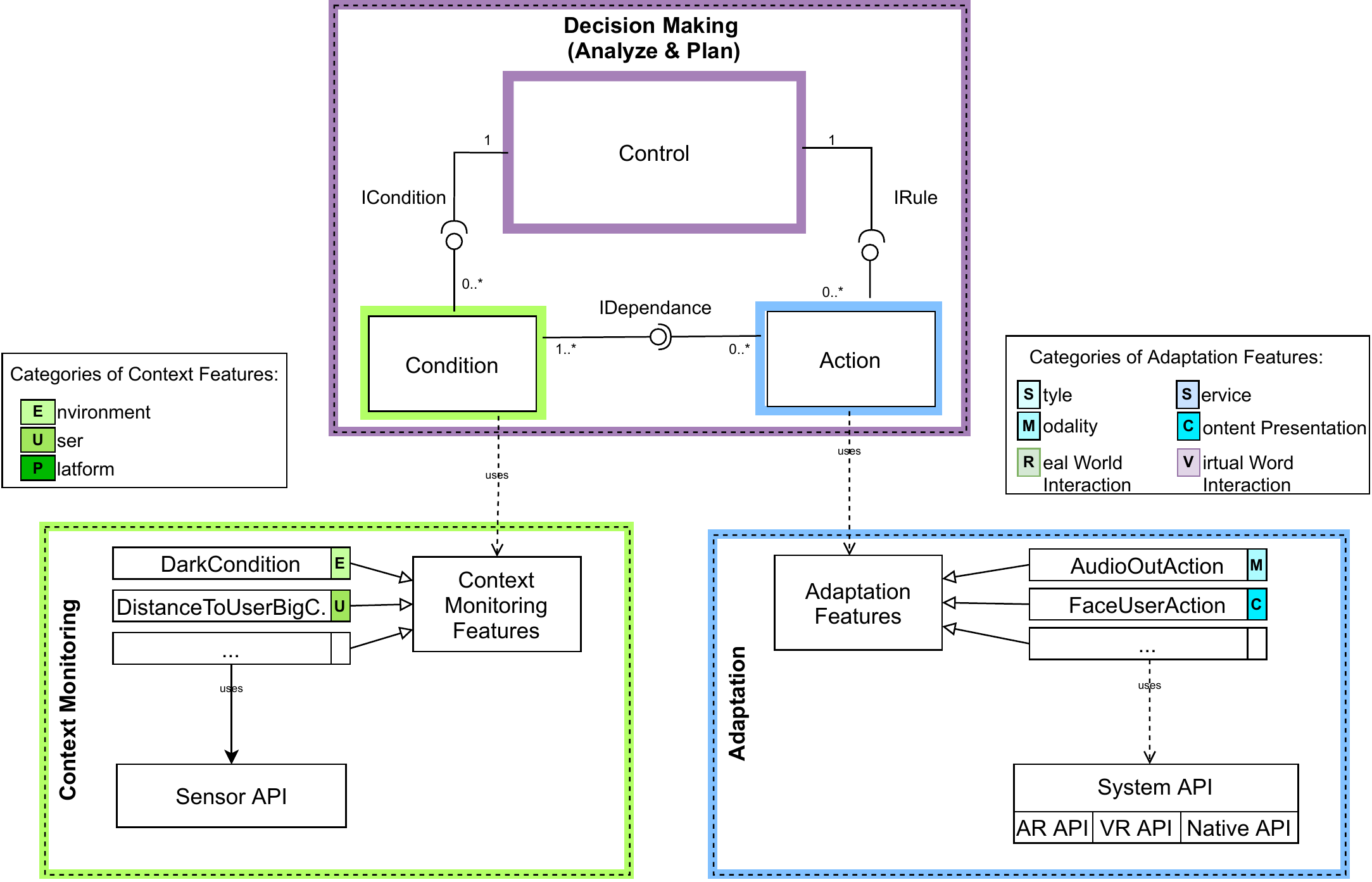}
    \caption{\textit{Monitoring}, \textit{Decision Making}, and \textit{Adaptation} in MAVAR}
  	\label{fig:solution_concept}
\end{figure}

A refined architectural overview of our monitoring and adaptation framework for virtual and augmented reality (MAVAR) based SADAS is depicted in~\autoref{fig:solution_concept}. It consists of three main components: \textit{Context Monitoring}, \textit{Decision  Making}, and \textit{Adaptation}.

The \textit{Context Monitoring} component is responsible for constantly collecting information about different kinds of context to enable the framework to react to them appropriately. All the information on the context is measured by sensors; partially real sensors like the camera or the inertial sensors like gyroscopes, partially in a more figurative sense like measuring information about the app such as positioning of virtual objects or usage. The sensor data is read out through \textit{Sensor API}s and used by several sub-components which are each responsible for checking on one specific context feature (e.g. \textit{DarkCondition} and \textit{DistanceToUserBigCondition} in~\autoref{fig:solution_concept}). The entirety of the \textit{Context Monitoring Features} is used by the \textit{Condition} and each of the context features has its own component responsible for monitoring it. The context observed by the \textit{Condition} sub-classes is categorized into \textit{Environment}, \textit{User}, and \textit{Platform Context}. The \textit{Environment Context} includes everything that impacts the system from the outside, such as noise or objects in the real world (in AR) or virtual environmental information (in VR). \textit{User Context} denotes any information available about the user, such as age or experience, but also the social context the user is currently in. Any information about the platform on which the system is running, like the availability of sensors or the compatibility with different software kits, is summarized in the \textit{Platform Context}.\\

The \textit{Decision Making} is led by the \textit{Control} component. This component supervises the active conditions and rules from the \textit{Context Monitoring} and \textit{Adaptation} components. The connections between the \textit{Control}, \textit{Rule}, and \textit{Condition} component make it possible for the components to work together closely and to incorporate and connect the \textit{Context Monitoring} and \textit{Adaptation} component.

The \textit{Adaptation} component is responsible for the execution of actions in response to a captured context change. The \textit{Adaptation Features} component consists of several sub-components (like the \textit{AudioOutRule} or the \textit{FaceUserRule} in~\autoref{fig:solution_concept}) specified to each execute a respective adaptation. To do so, the sub-components make use of different parts of the \textit{System API} to access the needed functions. Some of these APIs are for example for \textit{AR API}, which is necessary to influence the AR part of the application, or the \textit{Native API}, which is used to influence native system functions as the language. The adaptation features, which are executed by the \textit{Rule} sub-classes, are divided into the \textit{Style}, \textit{Modality}, \textit{Service}, \textit{Content Presentation}, \textit{Real-World}, and \textit{Virtual-World} changes. The \textit{Style} adaptation operation changes the look or behavior of single elements, the \textit{Modality} operation adjusts the sensory input and output the user utilizes to interact with the application, and \textit{Content Presentation} treats the way contents are presented to the user on the screen. Furthermore, the \textit{Service} change operation describes changes made on the device level regarding the type of device or features it offers, the \textit{Real-World} changes treat actions that are tied to objects from the real, physical surroundings of the device and the \textit{Virtual-World} changes describe adjustments of virtual objects either in an AR or VR scene.

To achieve the required functionality of monitoring and adaptation of DAS, it is necessary to connect adaptations to the context changes that should trigger them. In MAVAR, this is done by adding one or more conditions to a rule, which will react to changes in the context features monitored by the conditions with the adaptation which is implemented for it. To be more specific, a rule will become active and get executed as soon as all of its conditions are fulfilled at the same time. For cleanup purposes, there is also an \textit{unexecute} method, which will be called when one or more of a rule's conditions are not fulfilled anymore (making the rule inactive) and which is supposed to be used to reverse any effects of the rules execution that should only be active as long as its conditions are true.

The constant monitoring of the context through the conditions and the prompt execution of the adaptations through the rules is ensured by the control component. The control component acts as an observer for all of the condition and rule components.
%, which, in turn, inherits from the observable component.% so the control component can use the observer pattern to get informed on all changes in the conditions and rules. \\ 
Conditions report a change to the control when they detect a change in the context feature which they monitor, therefore they report either on it newly being true or newly being false. Rules report a change to the control when they are either executed (thereby executing an adaptation) or unexecuted (reversing an adaptation). As the rules depend on their registered conditions for changing, they report to the control if either all their conditions are true and were not all true before (rule gets executed), or if all conditions were true before and at least one newly turned false (rule gets unexecuted).

To make sure the context changes are detected, the control component regularly updates itself. If the update method is called, each of the registered rules is evaluated and in turn evaluates its respective conditions to check whether a context change has happened since the last evaluation and returns the new state of the context to the rule. If a change occurred, the condition also notifies the observing control component, causing a new update process. The rules receive the result of each of their respective conditions and react accordingly (for example by being calling their execute method). If a rule is executed or unexecuted, it notifies the observing control component, so a new update will be executed in case any context features were impacted by the rule's actions.

\section{Case Studies}
\label{sec:caseStudies}

In this section, two case studies are presented which show the benefit of our monitoring and adaptation framework for digital assistance systems. The first case study deals with an AR-based SADAS for maintenance scenarios. The second case study shows an example of a VR-based SADAS which supports warehouse management training in a virtual environment.

\subsection{Example 1: AR-based Context-aware Assistance for Maintenance Tasks}
\label{sec:study1}

As an example application of our monitoring and adaptation framework for DAS, a multi-platform and context-aware AR app for printer maintenance was developed. The app guides its user through the process of exchanging the ink cartridges of a printer step by step, with each step being described in a text window and illustrated by 3D arrows and other elements arranged on the printer.

Some of the monitoring and adaptation features are illustrated with screenshots in the following figures, with the left image illustrating the state before the adaptation and the right image showing it after adapting.

\begin{figure}[h]
  \centering
	\includegraphics[width=1.05\columnwidth]{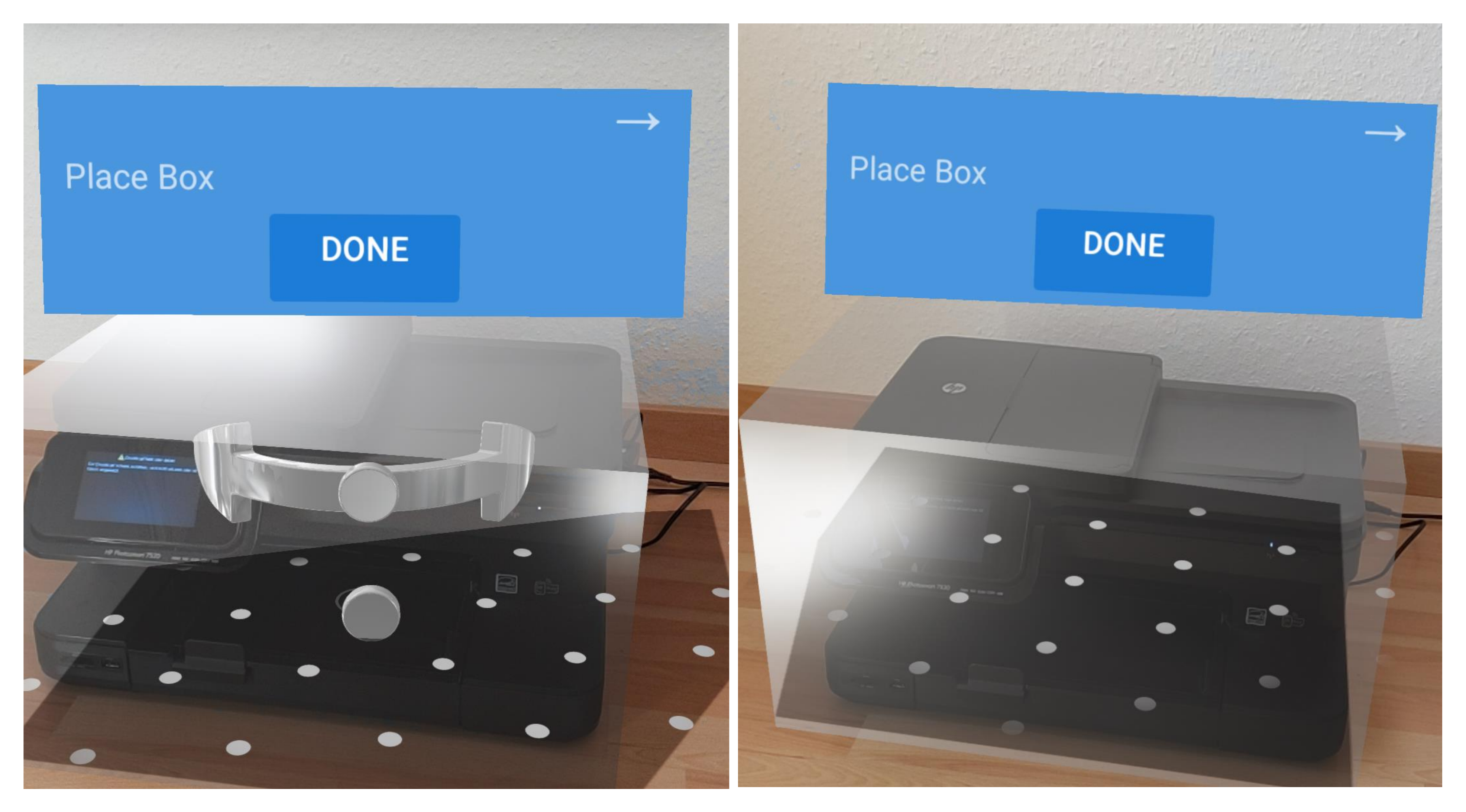}
    \caption{Experience-level adaptation: Instructions on how to turn elements are only shown to new users.}
  	\label{fig:case1-print1}
\end{figure}

In~\autoref{fig:case1-print1}, the effect of an adaptation responding to the user's experience is shown. It displays example control elements to illustrate how to do different transformations on 3D objects and thereby shows the user how they can, for example, rotate objects. As the action responsible for displaying the illustrations is connected to a condition monitoring the number of app uses, it is only executed on the first five uses of the app, so the user can use the app undisturbed once they got used to the controls.

\begin{figure}[h]
  \centering
	\includegraphics[width=1.05\columnwidth]{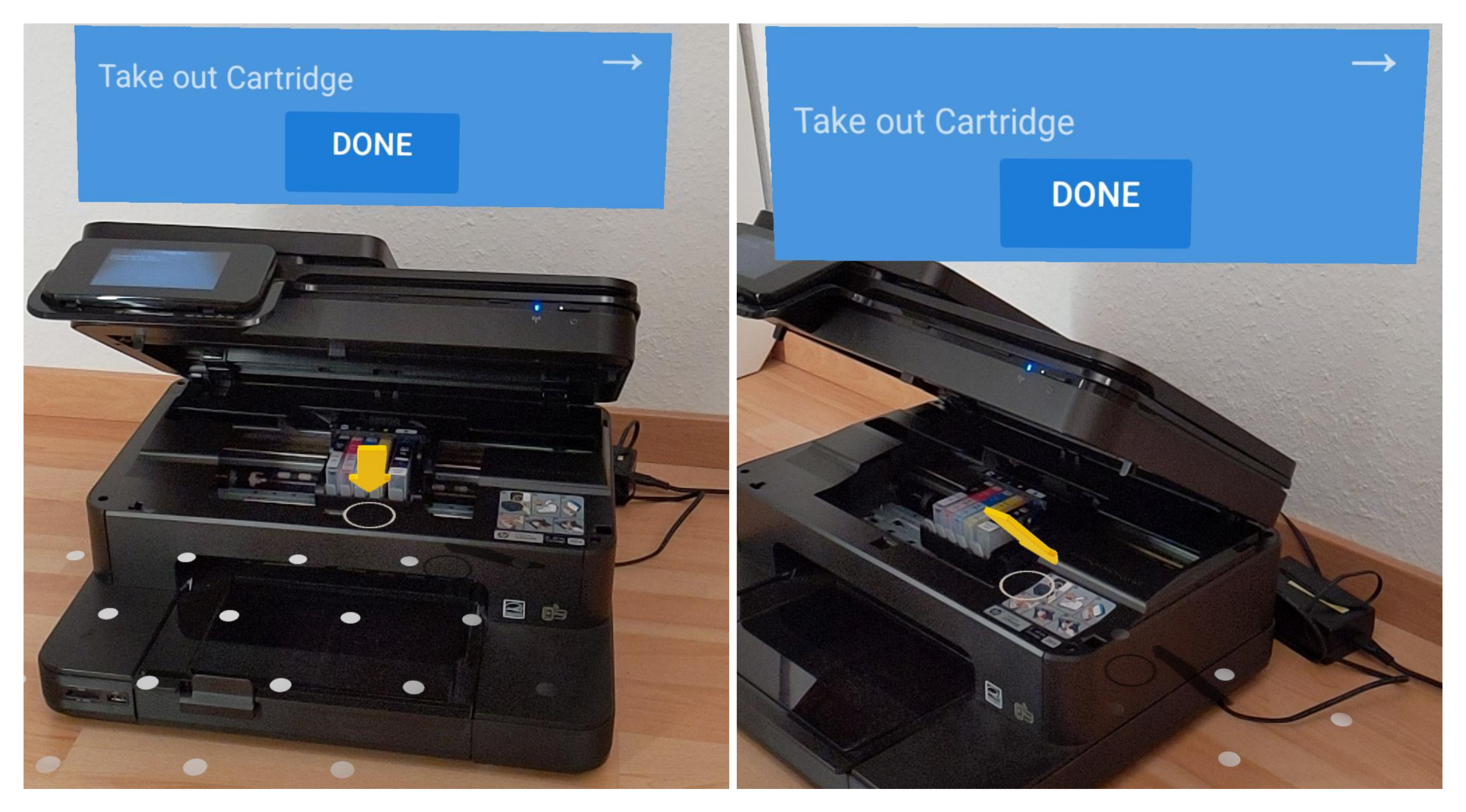}
    \caption{View-angle adaptation: The window is dynamically rotating to face the user regardless of their position.}
  	\label{fig:case1-print2}
\end{figure}

While working on the printer, the user has to access the printer from several angles. This can cause them to look at the message with the instructions from a very steep angle, which makes it hard or impossible to read. Of course, the user could move away from the printer, read the message, and then go back to the printer again, but that would be rather inconvenient and disruptive for the workflow. For this reason, the application uses an action that rotates the specified object, in this case, the message window, towards the user at all times (see~\autoref{fig:case1-print2}) to make sure it is always readable.

\begin{figure}[h]
  \centering
	\includegraphics[width=0.9\columnwidth]{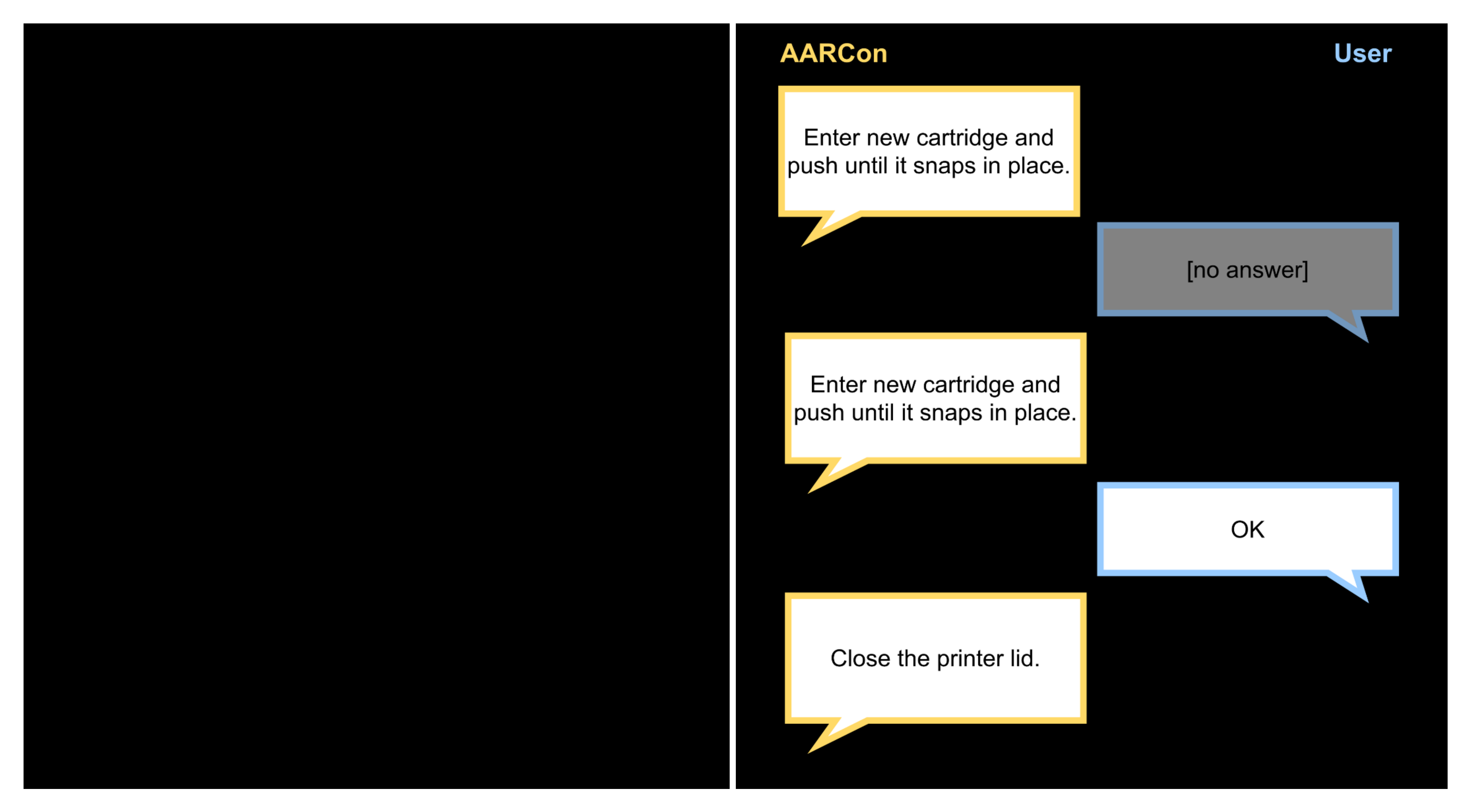}
    \caption{Modality adaptation: The application switches to a conversational UI if the AR camera is not available.}
  	\label{fig:case1-print3}
\end{figure}

\begin{figure}[h!]
  \centering
	\includegraphics[width=0.9\columnwidth]{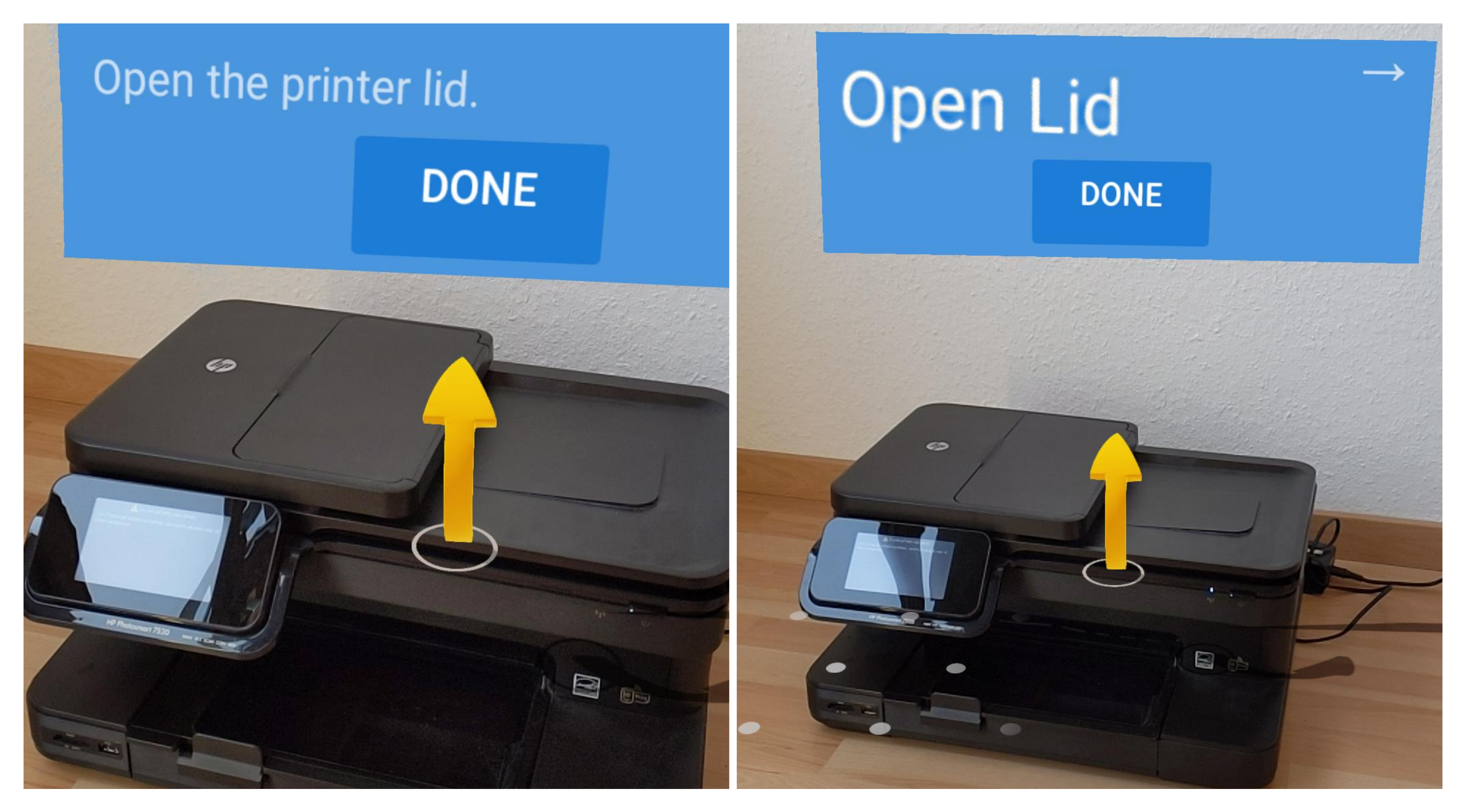}
    \caption{Distance-based adaptation: When the user moves away from the printer, the level of detail is decreased while the text size is increased.}
  	\label{fig:case1-print4}
\end{figure}

There is also a condition that turns true whenever the camera receives very little light input, which is usually the case when the device was laid down, for example, if the user needs their hands free. As this prevents the user from seeing any objects of the AR application, a voice interface is activated (see~\autoref{fig:case1-print3}).
It makes sure the description of the current task is read out to the user and it enables them to interact with the application using voice commands. This allows the user to interact with the application even if they are currently not able to hold the device to use it in AR mode.

\begin{figure}[h]
  \centering
	\includegraphics[width=1\columnwidth]{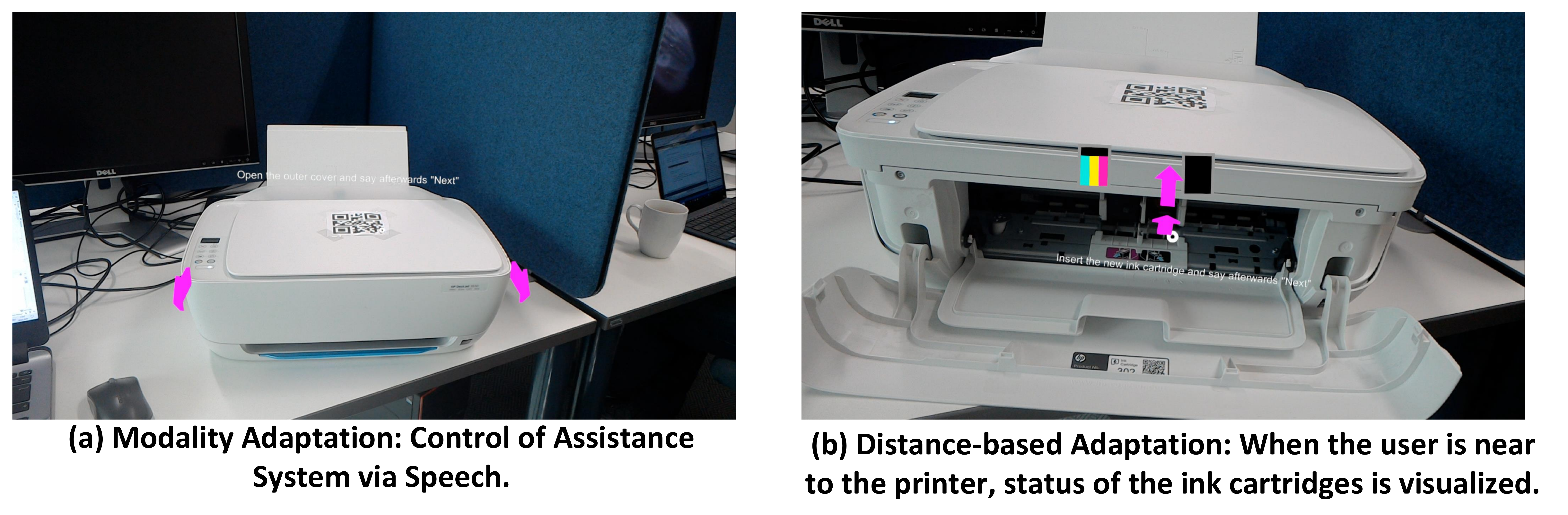}
    \caption{Context-aware AR Printer Maintenance App on HoloLens}
  	\label{fig:case1}
\end{figure}

~\autoref{fig:case1-print4} shows a change in the instruction window's level of detail. This is achieved using the action for this in combination with a condition that reacts to the user's distance to the printer, so the detail is lowered if the user is more than for example 1.2 meters away from the printer. Adjusting the level of detail can help the user to focus on the currently important task. It can also be used to make the user come closer to important objects or to create a simpler and tidier AR environment by removing information that is currently unnecessary.

These features, amongst others, aim to enhance the user's experience using the printer maintenance application. As they are all executed automatically based on context information, the user does not have to do anything to get an application that is at all times customized to the current situation.

Furthermore, in~\autoref{fig:case1}, screenshots from the same digital assistance system application are shown for a different target platform. Instead of an Android-based AR app like shown before, we now have the same app running on the HoloLens with the same context-awareness and UI adaptation features supported for the new target platform. 

In summary, the implementation of the above described AR-based digital assistance system shows how our MAVAR framework supports the monitoring and adaptation process of a DAS on different target platforms.

\subsection{Example 2: VR-based Context-aware Assistance for Warehouse Management Training}
\label{sec:study2}

\begin{figure}[h]
  \centering
	\includegraphics[width=1\columnwidth]{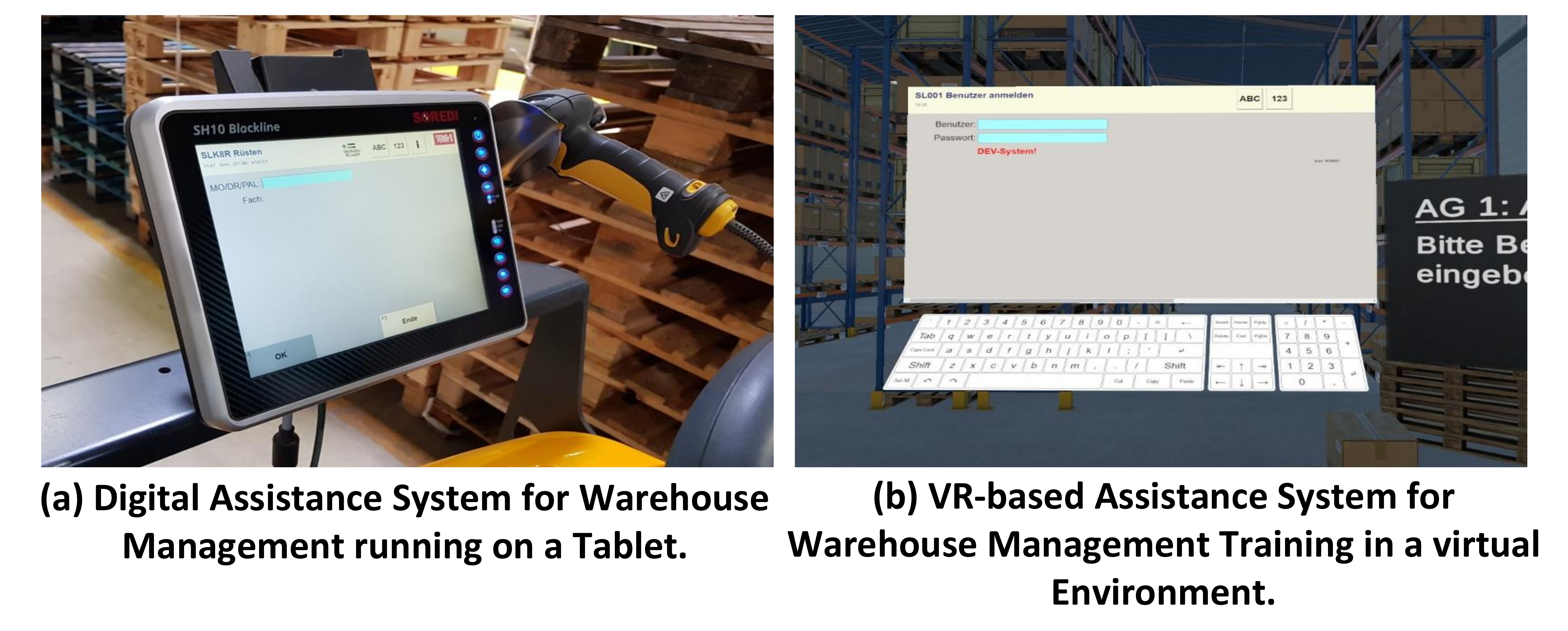}
    \caption{Warehouse Management}
  	\label{fig:case2-z1}
\end{figure}

As a further application scenario for our monitoring and adaptation framework for DAS, we present an example from the logistics domain. A typical task in this domain is warehouse management where employees have to pursue pick and order operations. As shown in~\autoref{fig:case2-z1}(a), a digital assistance system is used for supporting the employees in their tasks. In most cases, the picking process is paper-based in a classical sense, or sometimes there is a digital assistance system in the form of an application running on a tablet. In both cases, still, logistics executives often detect stock discrepancies and misplaced wares. This is due to the different levels of expertise of employees with the warehouse management system and order picking process. 

In addition, the differences in performance between order pickers are great. Some pick quickly and precisely, others with errors or even damage wares. Furthermore, in a real physical setting, training of the employees is expensive and time-consuming. It is often also the case that companies do not have a spare warehouse where they can train new employees and relocate moved wares to the original position after the training. Furthermore, it is desirable to offer repeatable and comparable training, especially for new or short-term staff.

\begin{figure}[h]
  \centering
	\includegraphics[width=1.05\columnwidth]{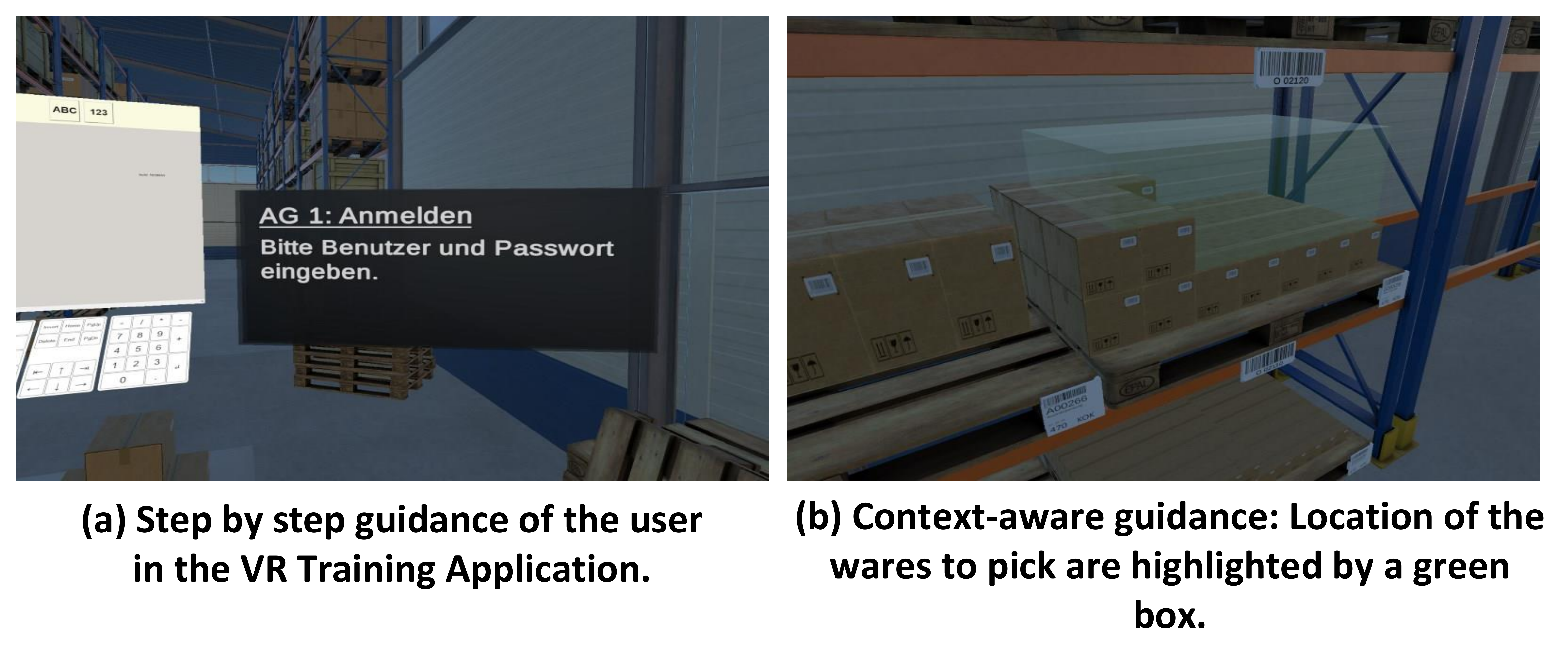}
    \caption{VR Warehouse Management Training Application}
  	\label{fig:case2-z2}
\end{figure}

To overcome these issues, we have developed a VR-based assistance system to support the training of the picking process. As shown in~\autoref{fig:case2-z1}(b), the same warehouse management application running on the tablet is provided in the virtual training environment. In addition to that, the core application logic of the warehouse management app can be further augmented through virtual elements that can be displayed in the virtual environment. This is, for example, used to realize a step by the guidance of the user of the VR training application. In~\autoref{fig:case2-z2}(a), one can see how additional textual descriptions help the user to accomplish the task. Similar to the AR-based assistance system, we have also here monitoring and adaptation features to guide the learning process in the warehouse management training app in the most suitable way. In~\autoref{fig:case2-z2}(b), for example, it is shown how the location of the wares which the user has to pick are highlighted by a green box. Based on such situation-aware information, the users can be guided through step-by-step context-aware information so that the effect of learning can increase.   

Furthermore, our VR-based assistance system is designed in such a way that different workflows in the area of warehouse management (Single-order and Multi-order picking with different kinds of exceptions) can be supported. For this purpose, the different training workflows are specified based on a process model (in our case BPMN) which can be edited according to the needs of the VR training application.  

To sum up, the illustration of the above described VR-based digital assistance system for warehouse management training shows that our monitoring and adaptation framework is applicable for different kinds of digital assistance systems and flexible to cover various workflows in this area.

\subsection{Discussion}
\label{sec:discussion}

While the above-described case studies illustrate the benefit of our monitoring and adaptation framework, there is still room for improvement of such self-adaptive digital assistance systems so that they can find their way to industrial practice. With this regard, it has to be mentioned that the implementation of our framework currently is in a prototypical state where several improvements regarding visualization of the AR/VR interfaces can be done to increase the usability and user experience (UX) of the end-users. In this context, a usability study should be conducted to assess the usability and UX of the resulting DAS based on our framework. Besides that, the efficiency and effectiveness of our monitoring and adaptation framework should be analyzed to check its applicability and benefit in further domains beyond maintenance and training.

\section{Conclusion and Outlook}
\label{sec:conclusion}

As a consequence of ongoing digital transformation, new technologies are emerging which change the way we are working and communicating with humans and machines. In this context, digital assistance systems play a crucial role as they provide means for supporting human-to-human and human-to-machine interactions. Furthermore, such digital assistance systems can be used to provide instructions and technical support in the working process as well as for training purposes. 

In this book chapter, we argue that existing digital assistance systems are mostly created focusing on the “design for all” paradigm neglecting the situation-specific tasks, skills, preferences, or environments of an individual human worker. To overcome this issue, we first discuss the main challenges in developing self-adaptive AR/VR-based digital assistance systems. After that, we present a monitoring and adaptation framework for supporting self-adaptive AR/VR-based digital assistance systems for Work 4.0. Our framework supports context monitoring as well as UI adaptation for AR/VR-based digital assistance systems. The benefit of our framework is shown based on exemplary case studies from different domains, e.g. context-aware maintenance application in augmented reality or warehouse management training in virtual reality.

Although the presented framework makes a further step to support and improve working processes of humans in times of Industry 4.0, further research according to assistance systems has to be done to reach a better degree of acceptance. For reaching this, further improvements in the areas of hardware and display technology are required so that dynamic 3d objects and information can be visualized on smart glasses or similar wearables. Beyond that, holographic 3d displays are emerging which could lead to the future of holographic working environments. Furthermore, intelligent techniques are necessary to improve object detection at run time that is a key enabler in assisting humans in working processes. In its current state, the implemented adaptation process in our framework follows a rule-based approach. Further optimization of UI adaptations can be reached through extending the adaptation manager by machine learning algorithms. This way, log data (context information, previous adaptations, and user feedback) can be analyzed to learn the most suitable adaptations for future context-of-use situations. In general, a broader acceptance of AR/VR technologies needs to be reached so that these technologies can be used to augment human abilities and thus improve their cognitive and physical tasks. 

\bibliographystyle{splncs04}
\bibliography{references}

\begin{thebibliography}{10}
\providecommand{\url}[1]{\texttt{#1}}
\providecommand{\urlprefix}{URL }
\providecommand{\doi}[1]{https://doi.org/#1}

\bibitem{bonekamp2015consequences}
Bonekamp, L., Sure, M.: Consequences of industry 4.0 on human labour and work
  organisation. Journal of Business and Media Psychology  \textbf{6}(1),
  33--40 (2015)

\bibitem{de2018work}
De~Vos, M.: Work 4.0 and the future of labour law. Available at SSRN 3217834
  (2018)

\bibitem{DBLP:conf/cdmake/FellmannRBMR17}
Fellmann, M., Robert, S., B{\"{u}}ttner, S., Mucha, H., R{\"{o}}cker, C.:
  Towards a framework for assistance systems to support work processes in smart
  factories. In: Holzinger, A., Kieseberg, P., Tjoa, A.M., Weippl, E.R. (eds.)
  Machine Learning and Knowledge Extraction - First {IFIP} {TC} 5, {WG} 8.4,
  8.9, 12.9 International Cross-Domain Conference, {CD-MAKE} 2017, Reggio di
  Calabria, Italy, August 29 - September 1, 2017, Proceedings. Lecture Notes in
  Computer Science, vol. 10410, pp. 59--68. Springer (2017).
  \doi{10.1007/978-3-319-66808-6\_5},
  \url{https://doi.org/10.1007/978-3-319-66808-6\_5}

\bibitem{DBLP:conf/hci/FischerSR018}
Fischer, H., Senft, B., Rittmeier, F., Sauer, S.: A canvas method to foster
  interdisciplinary discussions on digital assistance systems. In: Marcus, A.,
  Wang, W. (eds.) Design, User Experience, and Usability: Theory and Practice -
  7th International Conference, {DUXU} 2018, Held as Part of {HCI}
  International 2018, Las Vegas, NV, USA, July 15-20, 2018, Proceedings, Part
  {I}. Lecture Notes in Computer Science, vol. 10918, pp. 711--724. Springer
  (2018). \doi{10.1007/978-3-319-91797-9\_49},
  \url{https://doi.org/10.1007/978-3-319-91797-9\_49}

\bibitem{DBLP:conf/indin/GoreckySLZ14}
Gorecky, D., Schmitt, M., Loskyll, M., Z{\"{u}}hlke, D.:
  Human-machine-interaction in the industry 4.0 era. In: 12th {IEEE}
  International Conference on Industrial Informatics, {INDIN} 2014, Porto
  Alegre, RS, Brazil, July 27-30, 2014. pp. 289--294. {IEEE} (2014).
  \doi{10.1109/INDIN.2014.6945523},
  \url{https://doi.org/10.1109/INDIN.2014.6945523}

\bibitem{DBLP:conf/hcse/GottschalkYSE20}
Gottschalk, S., Yigitbas, E., Schmidt, E., Engels, G.: Model-based product
  configuration in augmented reality applications. In: Bernhaupt, R., Ardito,
  C., Sauer, S. (eds.) Human-Centered Software Engineering - 8th {IFIP} {WG}
  13.2 International Working Conference, {HCSE} 2020, Eindhoven, The
  Netherlands, November 30 - December 2, 2020, Proceedings. Lecture Notes in
  Computer Science, vol. 12481, pp. 84--104. Springer (2020).
  \doi{10.1007/978-3-030-64266-2\_5},
  \url{https://doi.org/10.1007/978-3-030-64266-2\_5}

\bibitem{DBLP:conf/hcse/GottschalkYSE20a}
Gottschalk, S., Yigitbas, E., Schmidt, E., Engels, G.: Proconar: {A} tool
  support for model-based {AR} product configuration. In: Bernhaupt, R.,
  Ardito, C., Sauer, S. (eds.) Human-Centered Software Engineering - 8th {IFIP}
  {WG} 13.2 International Working Conference, {HCSE} 2020, Eindhoven, The
  Netherlands, November 30 - December 2, 2020, Proceedings. Lecture Notes in
  Computer Science, vol. 12481, pp. 207--215. Springer (2020).
  \doi{10.1007/978-3-030-64266-2\_14},
  \url{https://doi.org/10.1007/978-3-030-64266-2\_14}

\bibitem{GrubertLZR17}
Grubert, J., et~al.: Towards pervasive augmented reality: Context-awareness in
  augmented reality. {IEEE} Trans. Vis. Comput. Graph.  \textbf{23}(6),
  1706--1724 (2017)

\bibitem{hinrichsen2018digital}
Hinrichsen, S., Bendzioch, S.: How digital assistance systems improve work
  productivity in assembly. In: International Conference on Applied Human
  Factors and Ergonomics. pp. 332--342. Springer (2018)

\bibitem{hold2017planning}
Hold, P., Erol, S., Reisinger, G., Sihn, W.: Planning and evaluation of digital
  assistance systems. Procedia Manufacturing  \textbf{9},  143--150 (2017)

\bibitem{hong2006new}
Hong, D., Shin, C., Oh, S., Woo, W.: A new paradigm for user interaction in
  ubiquitous computing environment. ISUVR 2006 pp. 41--44 (2006)

\bibitem{DBLP:conf/hci/JosifovskaYE19}
Josifovska, K., Yigitbas, E., Engels, G.: A digital twin-based multi-modal {UI}
  adaptation framework for assistance systems in industry 4.0. In: Kurosu, M.
  (ed.) Human-Computer Interaction. Design Practice in Contemporary Societies -
  Thematic Area, {HCI} 2019, Held as Part of the 21st {HCI} International
  Conference, {HCII} 2019, Orlando, FL, USA, July 26-31, 2019, Proceedings,
  Part {III}. Lecture Notes in Computer Science, vol. 11568, pp. 398--409.
  Springer (2019). \doi{10.1007/978-3-030-22636-7\_30},
  \url{https://doi.org/10.1007/978-3-030-22636-7\_30}

\bibitem{DBLP:conf/hcse/JovanovikjY0E20}
Jovanovikj, I., Yigitbas, E., Sauer, S., Engels, G.: Augmented and virtual
  reality object repository for rapid prototyping. In: Bernhaupt, R., Ardito,
  C., Sauer, S. (eds.) Human-Centered Software Engineering - 8th {IFIP} {WG}
  13.2 International Working Conference, {HCSE} 2020, Eindhoven, The
  Netherlands, November 30 - December 2, 2020, Proceedings. Lecture Notes in
  Computer Science, vol. 12481, pp. 216--224. Springer (2020).
  \doi{10.1007/978-3-030-64266-2\_15},
  \url{https://doi.org/10.1007/978-3-030-64266-2\_15}

\bibitem{keller2019benefit}
Keller, T., Bayer, C., Bausch, P., Metternich, J.: Benefit evaluation of
  digital assistance systems for assembly workstations. Procedia CIRP
  \textbf{81},  441--446 (2019)

\bibitem{DBLP:journals/computer/KephartC03}
Kephart, J.O., Chess, D.M.: The vision of autonomic computing. Computer
  \textbf{36}(1),  41--50 (2003). \doi{10.1109/MC.2003.1160055},
  \url{https://doi.org/10.1109/MC.2003.1160055}

\bibitem{DBLP:conf/ml4cps/KovacsAGUGS18}
Kovacs, K., Ansari, F., Geisert, C., Uhlmann, E., Glawar, R., Sihn, W.: A
  process model for enhancing digital assistance in knowledge-based
  maintenance. In: Beyerer, J., K{\"{u}}hnert, C., Niggemann, O. (eds.) Machine
  Learning for Cyber Physical Systems, Selected papers from the International
  Conference {ML4CPS} 2018, Karlsruhe, Germany, October 23-24, 2018. pp.
  87--96. Springer (2018). \doi{10.1007/978-3-662-58485-9\_10},
  \url{https://doi.org/10.1007/978-3-662-58485-9\_10}

\bibitem{DBLP:conf/eics/KringsYJ0E20}
Krings, S., Yigitbas, E., Jovanovikj, I., Sauer, S., Engels, G.: Development
  framework for context-aware augmented reality applications. In: Bowen, J.,
  Vanderdonckt, J., Winckler, M. (eds.) {EICS} '20: {ACM} {SIGCHI} Symposium on
  Engineering Interactive Computing Systems, Sophia Antipolis, France, June
  23-26, 2020. pp. 9:1--9:6. {ACM} (2020). \doi{10.1145/3393672.3398640},
  \url{https://doi.org/10.1145/3393672.3398640}

\bibitem{laddaga2004self}
Laddaga, R., Robertson, P.: Self adaptive software: A position paper. In:
  SELF-STAR: International Workshop on Self-* Properties in Complex Information
  Systems. vol.~31, p.~19. Citeseer (2004)

\bibitem{lasi2014industry}
Lasi, H., Fettke, P., Kemper, H.G., Feld, T., Hoffmann, M.: Industry 4.0.
  Business \& information systems engineering  \textbf{6}(4),  239--242 (2014)

\bibitem{LindlbauerFH19}
Lindlbauer, D., Feit, A.M., Hilliges, O.: Context-aware online adaptation of
  mixed reality interfaces. In: Proceedings of the 32nd Annual {ACM} Symposium
  on User Interface Software and Technology, {UIST} 2019, New Orleans, LA, USA,
  October 20-23, 2019. pp. 147--160 (2019)

\bibitem{DBLP:conf/icit2/NellesKMS16}
Nelles, J., Kuz, S., Mertens, A., Schlick, C.M.: Human-centered design of
  assistance systems for production planning and control: The role of the human
  in industry 4.0. In: {IEEE} International Conference on Industrial
  Technology, {ICIT} 2016, Taipei, Taiwan, March 14-17, 2016. pp. 2099--2104.
  {IEEE} (2016). \doi{10.1109/ICIT.2016.7475093},
  \url{https://doi.org/10.1109/ICIT.2016.7475093}

\bibitem{nikolenko2019digital}
Nikolenko, A., Sehr, P., Hinrichsen, S., Bendzioch, S.: Digital assembly
  assistance systems--a case study. In: International Conference on Applied
  Human Factors and Ergonomics. pp. 24--33. Springer (2019)

\bibitem{camar}
Oh, S., Woo, W., et~al.: Camar: Context-aware mobile augmented reality in smart
  space. Proc. of IWUVR  \textbf{9},  48--51 (2009)

\bibitem{DBLP:conf/etfa/Paelke14}
Paelke, V.: Augmented reality in the smart factory: Supporting workers in an
  industry 4.0. environment. In: Grau, A., Mart{\'{\i}}nez, H. (eds.)
  Proceedings of the 2014 {IEEE} Emerging Technology and Factory Automation,
  {ETFA} 2014, Barcelona, Spain, September 16-19, 2014. pp.~1--4. {IEEE}
  (2014). \doi{10.1109/ETFA.2014.7005252},
  \url{https://doi.org/10.1109/ETFA.2014.7005252}

\bibitem{russmann2015industry}
R{\"u}{\ss}mann, M., Lorenz, M., Gerbert, P., Waldner, M., Justus, J., Engel,
  P., Harnisch, M.: Industry 4.0: The future of productivity and growth in
  manufacturing industries. Boston Consulting Group  \textbf{9}(1),  54--89
  (2015)

\bibitem{salimi2015work}
Salimi, M.: Work 4.0: An enormous potential for economic growth in germany.
  ADAPT Bulletin  \textbf{16} (2015)

\bibitem{xu2018industry}
Xu, L.D., Xu, E.L., Li, L.: Industry 4.0: state of the art and future trends.
  International Journal of Production Research  \textbf{56}(8),  2941--2962
  (2018)

\bibitem{YigitbasS0E17}
Yigitbas, E., et~al.: Self-adaptive {UIs}: Integrated model-driven development
  of {UIs} and their adaptations. In: Proc. of the {ECMFA} 2017. pp. 126--141
  (2017)

\bibitem{DBLP:conf/hcse/YigitbasAJK0E18}
Yigitbas, E., Anjorin, A., Jovanovikj, I., Kern, T., Sauer, S., Engels, G.:
  Usability evaluation of model-driven cross-device web user interfaces. In:
  Bogdan, C., Kuusinen, K., L{\'{a}}rusd{\'{o}}ttir, M.K., Palanque, P.A.,
  Winckler, M. (eds.) Human-Centered Software Engineering - 7th {IFIP} {WG}
  13.2 International Working Conference, {HCSE} 2018, Sophia Antipolis, France,
  September 3-5, 2018, Revised Selected Papers. Lecture Notes in Computer
  Science, vol. 11262, pp. 231--247. Springer (2018).
  \doi{10.1007/978-3-030-05909-5\_14},
  \url{https://doi.org/10.1007/978-3-030-05909-5\_14}

\bibitem{DBLP:journals/corr/abs-2107-12772}
Yigitbas, E., Gorissen, S., Weidmann, N., Engels, G.: Collaborative software
  modeling in virtual reality. CoRR  \textbf{abs/2107.12772} (2021),
  \url{https://arxiv.org/abs/2107.12772}

\bibitem{DBLP:conf/mc/YigitbasHE19}
Yigitbas, E., Heind{\"{o}}rfer, J., Engels, G.: A context-aware virtual reality
  first aid training application. In: Alt, F., Bulling, A., D{\"{o}}ring, T.
  (eds.) Proc. of Mensch und Computer 2019. pp. 885--888. {GI} / {ACM} (2019)

\bibitem{DBLP:journals/pacmhci/YigitbasHRASE19}
Yigitbas, E., Hottung, A., Rojas, S.M., Anjorin, A., Sauer, S., Engels, G.:
  Context- and data-driven satisfaction analysis of user interface adaptations
  based on instant user feedback. Proc. {ACM} Hum. Comput. Interact.
  \textbf{3}({EICS}),  19:1--19:20 (2019). \doi{10.1145/3331161},
  \url{https://doi.org/10.1145/3331161}

\bibitem{YigitbasJJKAE19}
Yigitbas, E., Josifovska, K., Jovanovikj, I., Kalinci, F., Anjorin, A., Engels,
  G.: Component-based development of adaptive user interfaces. In: Proceedings
  of the {ACM} {SIGCHI} Symposium on Engineering Interactive Computing Systems,
  {EICS} 2019, Valencia, Spain, June 18-21, 2019. pp. 13:1--13:7 (2019)

\bibitem{DBLP:conf/interact/YigitbasJE21}
Yigitbas, E., Jovanovikj, I., Engels, G.: Simplifying robot programming using
  augmented reality and end-user development. In: Ardito, C., Lanzilotti, R.,
  Malizia, A., Petrie, H., Piccinno, A., Desolda, G., Inkpen, K. (eds.)
  Human-Computer Interaction - {INTERACT} 2021 - 18th {IFIP} {TC} 13
  International Conference, Bari, Italy, August 30 - September 3, 2021,
  Proceedings, Part {I}. Lecture Notes in Computer Science, vol. 12932, pp.
  631--651. Springer (2021). \doi{10.1007/978-3-030-85623-6\_36},
  \url{https://doi.org/10.1007/978-3-030-85623-6\_36}

\bibitem{DBLP:conf/interact/YigitbasJ0E19}
Yigitbas, E., Jovanovikj, I., Sauer, S., Engels, G.: On the development of
  context-aware augmented reality applications. In: Abdelnour{-}Nocera, J.L.,
  Parmaxi, A., Winckler, M., Loizides, F., Ardito, C., Bhutkar, G., Dannenmann,
  P. (eds.) Beyond Interactions - {INTERACT} 2019 {IFIP} {TC} 13 Workshops,
  Paphos, Cyprus, September 2-6, 2019, Revised Selected Papers. Lecture Notes
  in Computer Science, vol. 11930, pp. 107--120. Springer (2019).
  \doi{10.1007/978-3-030-46540-7\_11},
  \url{https://doi.org/10.1007/978-3-030-46540-7\_11}

\bibitem{DBLP:conf/vrst/YigitbasJSE20}
Yigitbas, E., Jovanovikj, I., Scholand, J., Engels, G.: {VR} training for
  warehouse management. In: Teather, R.J., Joslin, C., Stuerzlinger, W.,
  Figueroa, P., Hu, Y., Batmaz, A.U., Lee, W., Ortega, F.R. (eds.) {VRST} '20:
  26th {ACM} Symposium on Virtual Reality Software and Technology. pp.
  78:1--78:3. {ACM} (2020)

\bibitem{DBLP:conf/seams/YigitbasKJE21}
Yigitbas, E., Karakaya, K., Jovanovikj, I., Engels, G.: Enhancing
  human-in-the-loop adaptive systems through digital twins and {VR} interfaces.
  In: 16th International Symposium on Software Engineering for Adaptive and
  Self-Managing Systems, SEAMS@ICSE 2021, Madrid, Spain, May 18-24, 2021. pp.
  30--40. {IEEE} (2021). \doi{10.1109/SEAMS51251.2021.00015},
  \url{https://doi.org/10.1109/SEAMS51251.2021.00015}

\bibitem{DBLP:journals/corr/abs-2107-00377}
Yigitbas, E., Klauke, J., Gottschalk, S., Engels, G.: {VREUD} - an end-user
  development tool to simplify the creation of interactive {VR} scenes. CoRR
  \textbf{abs/2107.00377} (2021), \url{https://arxiv.org/abs/2107.00377}

\bibitem{Yigitbas016}
Yigitbas, E., Sauer, S.: Engineering context-adaptive uis for task-continuous
  cross-channel applications. In: Human-Centered and Error-Resilient Systems
  Development - {IFIP} {WG} 13.2/13.5 Joint Working Conference. pp. 281--300
  (2016)

\bibitem{DBLP:conf/eics/EnesScaffolding}
Yigitbas, E., Sauer, S., Engels, G.: Using augmented reality for enhancing
  planning and measurements in the scaffolding business. In: {EICS} '21: {ACM}
  {SIGCHI} Symposium on Engineering Interactive Computing Systems, virtual,
  June 8-11, 2021. {ACM} (2021), \url{https://doi.org/10.1145/3459926.3464747}

\bibitem{DBLP:conf/mc/YigitbasTE20}
Yigitbas, E., Tejedor, C.B., Engels, G.: Experiencing and programming the
  {ENIAC} in {VR}. In: Alt, F., Schneegass, S., Hornecker, E. (eds.) Mensch und
  Computer 2020. pp. 505--506. {ACM} (2020)

\end{thebibliography}

\end{document}